\documentclass{article}

\usepackage{fancyhdr}

\usepackage{arxiv}

\usepackage[utf8]{inputenc} 
\usepackage[T1]{fontenc}    
\usepackage{hyperref}       
\usepackage{url}            
\usepackage{booktabs}       
\usepackage{amsfonts}       
\usepackage{nicefrac}       
\usepackage{microtype}      
\usepackage{lipsum}		
\usepackage{graphicx}
\usepackage{doi}
\usepackage{framed,multirow}
\usepackage{pifont}
\newcommand{\cmark}{\ding{51}}%
\newcommand{\xmark}{\ding{55}}%
\usepackage{float}
\usepackage{booktabs, makecell, tabularx}
\usepackage{multirow}

\usepackage{textcomp}

\usepackage{graphicx}
\usepackage{amsmath,amssymb}
\usepackage{float}
\usepackage{booktabs, makecell, tabularx}
\usepackage{rotating}
\usepackage[export]{adjustbox}
\usepackage{booktabs}

\usepackage{amsmath,amsfonts}
\usepackage{algorithmic}
\usepackage{algorithm}
\usepackage{array}
\usepackage[caption=false,font=normalsize,labelfont=sf,textfont=sf]{subfig}
\usepackage{textcomp}
\usepackage{stfloats}
\usepackage{url}
\usepackage{verbatim}
\usepackage{graphicx}
\usepackage{cite}
\title{D-TrAttUnet: Dual-Decoder Transformer-Based Attention Unet  Architecture for Binary and Multi-classes Covid-19 Infection Segmentation \thanks{This work has been submitted to the IEEE for possible publication. Copyright may be transferred without notice, after which this version may no longer be accessible.}}

\author{ \href{https://sciprofiles.com/profile/FaresBougourzi}{\includegraphics[scale=0.06]{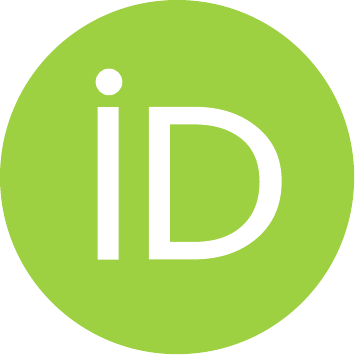}\hspace{1mm}Fares Bougourzi}
\\
 National Research Council of Italy, 73100 Lecce, Italy
	\\ and University Paris-Est Creteil, Laboratoire LISSI, 94400 \\Vitry sur Seine, Paris, France \\  \texttt{faresbougourzi@gmail.com} 
	\And
	\href{https://orcid.org/0000-0000-0000-0000}{\includegraphics[scale=0.06]{orcid.pdf}\hspace{1mm}Cosimo Distante} \\
	Institute of Applied Sciences and Intelligent\\ Systems, National Research Council of Italy,\\ 73100 Lecce, Italy \\ Department of Innovation Engineering, University \\of Salento, 73100 Lecce, Italy \\
	\texttt{cosimo.distante@cnr.it} \\	
	\And
	\href{https://orcid.org/0000-0000-0000-0000}{\includegraphics[scale=0.06]{orcid.pdf}\hspace{1mm} Fadi Dornaika} \\
	Ho Chi Minh City Open Univesity,\\
	   97 Vo Van Tan, Ward Vo Thi Sau, District 3,\\
    Ho Chi Minh City,  70000,   Vietnam   \\
	\texttt{fdornaika@gmail.com} \\  
	\And
	\href{https://orcid.org/0000-0000-0000-0000}{\includegraphics[scale=0.06]{orcid.pdf}\hspace{1mm}Abdelmalik Taleb-Ahmed} \\
	IEMN UMR CNRS 8520, Université \\Polytechnique Hauts de France, UPHF\\
	\texttt{Abdelmalik.Taleb-Ahmed@uphf.fr} \\
}
\renewcommand{\shorttitle}{ }

\hypersetup{
pdftitle={A template for the arxiv style},
pdfsubject={q-bio.NC, q-bio.QM},
pdfauthor={David S.~Hippocampus, Elias D.~Striatum},
pdfkeywords={First keyword, Second keyword, More},
}

\begin{document}
\maketitle

\begin{abstract}

In the last three years, the world has been facing a global crisis caused by Covid-19 pandemic. Medical imaging has been playing a crucial role in the fighting against this disease and saving the human lives. Indeed, CT-scans has proved their efficiency in diagnosing, detecting, and following-up the Covid-19 infection. In this paper, we propose a new Transformer-CNN based approach for Covid-19 infection segmentation from the CT slices. The proposed D-TrAttUnet architecture has an Encoder-Decoder structure, where compound Transformer-CNN encoder and Dual-Decoders are proposed. The Transformer-CNN encoder is built using Transformer layers, UpResBlocks, ResBlocks and max-pooling layers. The Dual-Decoder consists of two identical CNN decoders with attention gates. The two decoders are used to segment the infection and the lung regions simultaneously and the losses of the two tasks are joined. 

The proposed D-TrAttUnet architecture is evaluated for both Binary and Multi-classes Covid-19 infection segmentation. The experimental results prove the efficiency of the proposed approach to  deal with the complexity of Covid-19 segmentation task from limited data. Furthermore, D-TrAttUnet architecture outperforms three baseline CNN segmentation architectures (Unet, AttUnet and Unet++) and three state-of-the-art architectures (AnamNet, SCOATNet and CopleNet), in both Binary and Mutli-classes segmentation tasks.  

\end{abstract}



\section{Introduction}
\label{sec:introduction}

Since December 2019, the world has been in a global crisis due to the spread of Covid-19 disease. The first step in the fight against this pandemic is to detect and isolate the infected person to stop the infection cycle. In fact, RT-PCR test has been considered as the golden standard for detecting the infected individuals. However, this test has a considerable false-negative rate \cite{kucirka_variation_2020, bougourzi_recognition_2021}. On the other hand, RT-PCR test has very limited use for following up the patient's state and the progression of the disease \cite{jin_rapid_2020, wu_jcs_2021, bougourzi_ilc-unet_2022}. To deal with these issues medical imaging modalities have been widely used as major or supporting tool \cite{Vantaggiatoetal}.

CT-scans have been widely used for Covid-19 analysis. CT-scans are a very informative screening tool, thus, it can be used for Covid-19 detection \cite{jin_rapid_2020, wu_jcs_2021}, Segmentation \cite{bougourzi_ilc-unet_2022}, Severity prediction \cite{li_deep-learning-based_2021} and Covid-19 infection percentage estimation \cite{bougourzi_per-covid-19_2021}. Covid-19 infection segmentation is very important to recognize the infection and following up the patient's state \cite{fan_inf-net_2020, zhao2021scoat, bougourzi2023pdatt}. However, developing an efficient machine learning approach to segment the infection is very challenging task \cite{wang_noise-robust_2020, paluru_anam-net_2021, wang_focus_2022}.

The main challenge in segmenting Covid-19 infections is the lack of an adequate data to train machine learning models \cite{elharrouss_encoderdecoder-based_2021, wang_focus_2022}, especially deep learning methods, which requires huge labelled data \cite{bougourzi_deep_2022, bougourzi_cnn_2022}. On the other hand, the infection has high variability in intensity, shape, position and type, which makes the segmentation task very challenging \cite{kumar_singh_lunginfseg_2021, Laradji_2021_WACV}. Covid-19 infection stage (early vs advanced), symptoms (asymptomatic vs symptomatic patients), and severity are additional factors that complicate the segmentation task \cite{sun_systematic_2020}. In more details, the infection usually appears as ground-glass opacification (GGO) especially in the early stages \cite{salehi_coronavirus_2020, sun_systematic_2020}. Generally, GGO has blurred edges and low contrast with the surrounding lung tissues \cite{salehi_coronavirus_2020, sun_systematic_2020}. 
In later stages, the infection usually appears as a mixture of GGO and consolidation. At these stages, the infection segmentation challenges are compound of the blurred edges and low contrast of GGO, and the resemblance between the consolidation, the lungs walls and the other lung tissues \cite{salehi_coronavirus_2020, sun_systematic_2020}.

In the literary, Covid-19 infection segmentation can be categorized as Binary segmentation (infection or non-infection)  \cite{fan_inf-net_2020, wang_noise-robust_2020, paluru_anam-net_2021, liu_covid-19_2021} or Multi-classes segmentation (non-infection, GGO or consolidation) \cite{zhao2021scoat, wang_focus_2022}. Binary segmentation is quantitative segmentation that shows how much the infection is spread in the lungs. On the other hand, the multi-classes segmentation provides information not only about the degree of infection spread, but it gives better idea about the stage, progress and severity of the infection \cite{zhao2021scoat, wang_focus_2022, bougourzi_ilc-unet_2022}. Despite, the big importance of segmenting different infection classes, only few works have addressed this task due to data limitation for multi-classes Covid-19 segmentation \cite{zhao2021scoat, wang_focus_2022}.

The aim of this work is exploit the recent strengths of both CNNs and Transformers to provide an efficient solution for Covid-19 infection segmentation as binary and Multi-classes segmentation tasks. The proposed approach has demonstrated its efficiency in extracting local context information, long-range dependencies, and global context information through the proposed CNN-Transformer encoder.
To guide the models to concentrate inside the lungs and discard non-lung tissues, Dual Decoders is proposed for segmenting the infection and the lungs simultaneously. In summary, the main contributions of this paper are:

\begin{itemize}

\item A hybrid CNN-Transformer network is proposed that leverages the strengths of Transformers and CNNs to extract high-level features during the encoding phase.

\item   The proposed D-TrAttUnet encoder consists of two paths; the Transformer path and the Unet-like Fusion path. The Transformer path considers 2D patches of the input image as input, and consecutive Transformer layers to extract high representations at different levels. Four different Transformer features at different levels are injected into the Unet-like Fusion Encoder through UpResBlocks. On the other hand, the first layer of the Unet-like path uses the convolutional layers on the input image. The following Unet-Like Fusion layers combine the Transformer features with the previous layer of the Unet-Like path through concatenation and ResBlocks.

\item   The proposed D-TrAttUnet decoder consists of dual identical decoders. The objective of using two decoders is to segment Covid-19 infection and the lung regions simultaneously. Each decoder has four Attention Gates (AG), ResBlocks and bilinear Upsampling  layers similar to the Attion Unet (AttUnet) architecture, taking advantage of CNN-Transformer and multi-task tricks.

\item To evaluate the performance of our proposed architecture, both binary infection segmentation and multi-classes infection segmentation are investigated using three publicly available datasets. 

\item The comparison with three baseline architectures (Unet \cite{ronneberger_u-net_2015}, Att-Unet \cite{oktay_attention_2018}, and Unet++ \cite{zhou_unet_2018}) and three state-of-the-art architectures for Covid-19 segmentation (CopleNet \cite{wang_noise-robust_2020}, AnamNet \cite{paluru_anam-net_2021}, and SCOATNet \cite{zhao2021scoat}), demonstrates the superiority of our proposed D-TrAttUnet architecture in both Binary and Multi-classes Segmentation tasks. The codes of the proposed D-TrAttUnet will be publicly available at. \url{https://github.com/faresbougourzi/D-TrAttUnet}. 
 
\end{itemize}

This paper is organised as follows: Section \ref{S:2} presents some related work on CNN-based and Transformer-based segmentation architectures and segmentation of Covid-19 infections from CT scans. In section \ref{S:3}, we describe our proposed approach. Section \ref{S:4} consists of the description of the datasets used and the evaluation metrics. Section \ref{S:5} presents and discusses the experiments and results. Section \ref{S:6} shows some segmentation examples. Finally, section \ref{S:7} concludes this paper.

\section{Related Work}
\label{S:2}

In this section, we will briefly describe the related works on three aspects: CNN-based segmentation architectures, Transformers in computer vision and some Covid-19 infection segmentation state-of-the-art approaches.

\subsection{CNN Segmentation Architectures}

Since the great success of the first deep  CNN architecture ``Alexnet'' \cite{krizhevsky_imagenet_2012} in ImageNet \cite{deng_imagenet_2009} challenge in 2012, CNNs have reached the state of the art performance in many computer vision and machine learning tasks \cite{bougourzi_deep_2022, bougourzi_cnn_2022}. Segmentation tasks have been influenced by the great success of the CNNs and therefore many CNN architectures have proved their ability to segment many complicated medical imaging tasks \cite{ronneberger_u-net_2015 ,zhou_evolutionary_2020, tomar_fanet_2022}.  Since  Unet architecture  \cite{ronneberger_u-net_2015} was proposed in 2017, great progress has been made and a lot of Unet variants have been proposed such as  Attenion Unet (Att-Unet) \cite{oktay_attention_2018}, Unet++ \cite{zhou_unet_2018}, ResUnet \cite{zhang_road_2018}.

Unet \cite{ronneberger_u-net_2015}  is a CNN architecture with Encoder-Decoder structure. Unet's encoder consists of  consecutive CNN layers. Each layer contains convolutional and mapooling layers. On the other hand, the decoder consists of  consecutive decovolutional layers. The encoder and decoder are connected by skip connections, where encoder feature maps are concatenated with the decoder features to maintain fine-grained details by passing them to the decoder. This forms the ``U-shape''. In  \cite{oktay_attention_2018}, O. Oktay et al. proposed Attention Gate (AT) to determine the salient regions by using the encoder and decoder feature maps simultaneously.




\subsection{Transformers in CV}
Transformers are capable of capturing long-range  dependencies between sequence elements. Therefore, Transformers are widely used in the Natural Language Processing (NLP) domain \cite{khan_transformers_2021}. Inspired by the great success in the NLP domain, transformers have also been extensively studied in the computer vision domain in the last two years. \cite{khan_transformers_2021}. Transformers have shown promising results in many computer vision tasks and many transformer-based architectures have been proposed such as ViT \cite{dosovitskiy_image_2020}, Swin Transformer \cite{liu_swin_2021}, and Deit \cite{pmlr-v139-touvron21a}.

Similarly, Transformers have got much interest in Medical imaging domain \cite{shamshad_transformers_2022}. Indeed, Transformers have shown promising performance in many medical imaging tasks such as classification \cite{dai_transmed_2021}, detection \cite{shen_cotr_2021} and segmentation \cite{hatamizadeh_unetr_2022}. Since the focus of this work is segmentation task,  some transformer-based segmentation approaches will be describes. The segmentation architectures can be classified as 2-D \cite{wu_fat-net_2022, petit_u-net_2021} or 3-D modalities \cite{hatamizadeh_unetr_2022, wang_transbts_2021}.


In \cite{wu_fat-net_2022},  H. Wu et al. proposed a CNN-Transformer architecture called ``Fat-Net'', where two encoders (CNN and Transformer encoders) are used. The feature maps of the two decoders were concatenated to have richer features from the two encoders.  The Squeeze and Excitation (SE) module \cite{hu_squeeze-and-excitation_2018} was applied on the concatenated features to  identify the most important feature correlations from different feature channels. Fat-Net was evaluated for skin lesion segmentation using four public datasets.   In the U-Transformer \cite{petit_u-net_2021} architecture, Multi-Head Self-Attention and  Cross Attention modules were injected into the Unet architecture. These two modules were placed at the skip connection to learn the global
context information from the Unet encoder and pass them to its decoder \cite{petit_u-net_2021}. The U-Transformer architecture performed well in two abdominal CT-image datasets \cite{petit_u-net_2021}. In  \cite{hatamizadeh_unetr_2022}, A. Hatamizadeh et al. proposed a transformer-based architecture for multi-classes 3D segmentation called ``UNETR''. The encoder of UNETR  was constructed by a transformer, from which  four levels features are obtained and rescaled by deconvolutional layers. The rescaled maps  were connected to the CNN decoder via skip connections at different resolutions, forming the ``U-shape''.

\subsection{Covid-19 infection Segmentation}

Since the emergence of the infectious respiratory virus SARS-CoV-2 at the end of 2019 (in Wuhan, China), numerous machine learning approaches have been proposed to combat this deadly disease. CT -Scans along with machine learning approaches have been widely used for diagnosis \cite{bougourzi_recognition_2021, harmon_artificial_2020, shamsi_uncertainty-aware_2021}, severity prediction \cite{li_deep-learning-based_2021}, and quantification of  covid-19 infection \cite{bougourzi_per-covid-19_2021, chen_hadcnet_2022}.


In Covid-19 analysis, segmentation of Covid-19 infections is one of the most studied tasks in the last three years \cite{wang_focus_2022}. Indeed, CNN-based architectures have achieved the state of the art performance in Covid-19 infection segmentation \cite{fan_inf-net_2020, wang_noise-robust_2020, zhao2021scoat, liu2021covid, paluru_anam-net_2021}. In \cite{wang_noise-robust_2020}, G. Wang et al. propose a segmentation architecture for segmenting COVID -19 infection lesions called ''COPLE-Net''.The COPLE-Net architecture is a Unet variant in which a combination of max-pooling and average-pooling is introduced in the encoding phase. Moreover, a bridge module was applied to the encoder output to improve its semantic representation and learn multiple lesion scale features. In addition, COPLE-Net was designed to learn from noisy data that were not labelled by radiologists.

AnamNet \cite{paluru_anam-net_2021} is a lightweight
CNN with 7.8 times fewer parameters than Unet. Therefore, AnamNet is easy to train and has low inference time which makes it suitable even for mobile devices. On the other hand, AnamNet has a Unet-like structure with a proposed AD-block inserted after the downsampling block of the encoder and the upsampling block of the decoder
In short, the AD -block  was built using one by one convolutional depthwise squeezing layer, followed by another three by three convolutional layer, and finally one by one convolutional layer was used for depthwise stretching. 
In \cite{zhao2021scoat}, S. Zhao et al. proposed a Unet++ \cite{zhou_unet_2018} variant architecture called ``SCOATNet'', in which a new spacial-wise and channel-wise attention  modules are proposed. The attention modules were inserted between the dense convolutional blocks of the skip pathways of the Unet++ architecture.

\section{The Proposed Approach}
\label{S:3}

\begin{figure}[htp]
    \centering
    \includegraphics[width =3.5in, height = 2.2in] {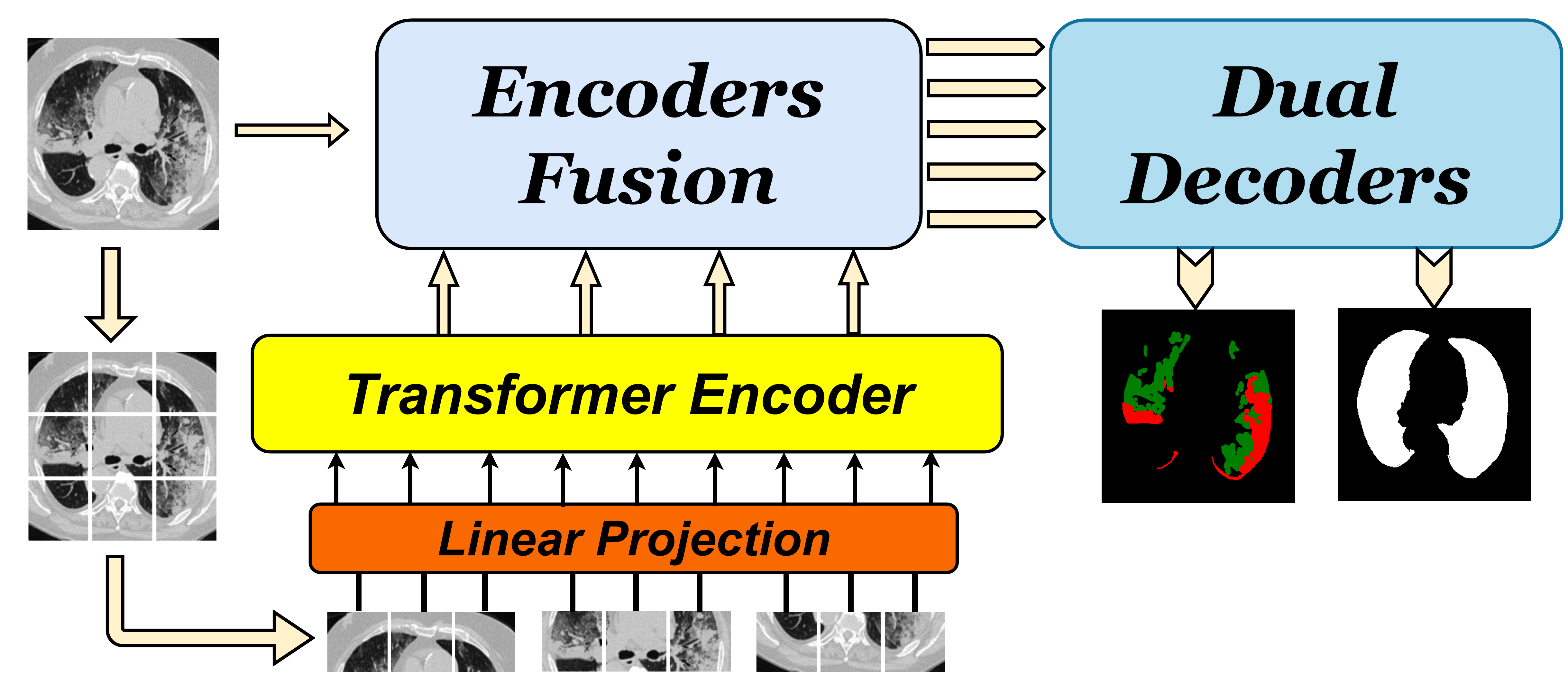} 
    \caption{The summary of our proposed D-TrAttUnet approach. }
    \label{fig:approachsum}
\end{figure}

\begin{figure}[htp]
    \includegraphics[width =6.6in, height = 4.6in] {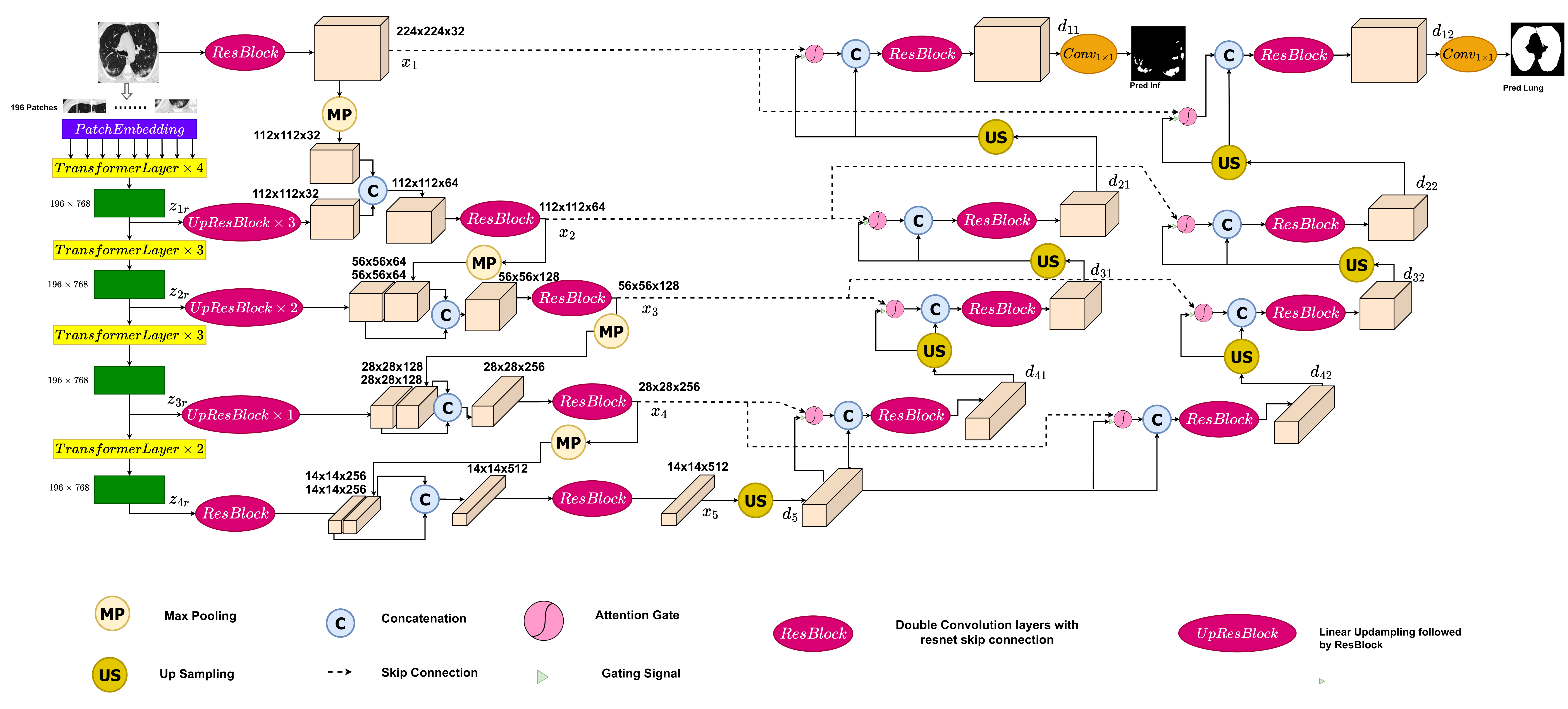} 
    \caption{Detailed Structure of the proposed D-TrAttUnet architecture. }
    \label{fig:approachA}
\end{figure}

The general structure of our proposed approach for Covid-19 segmentation is summarized in Figure \ref{fig:approachsum}.
The proposed D-TrAttUnet architecture is a compound CNN Transformer architecture with Unet-like shape and exploiting the Attention Gate (AG). Figure \ref{fig:approachA} shows the detailed structure of the proposed  D-TrAttUnet. The encoder of D-TrAttUnet exploits both the transformer layers and ResBlocks to extract rich features. Since Covid-19 infection problem has high variability in shape, size and position \cite{sun_systematic_2020, kumar_singh_lunginfseg_2021, Laradji_2021_WACV}, it is important to extract richer features from the input images. Our approach combines the extracted local features using CNN layers and  global  features using Transformer layers.

In our proposed D-TrAttUnet, the encoder has two paths: the Unet-like path and the Transformer path. The input image $x \in \mathbb{R}^{H\times W\times C}$, where $H$, $W$ and $C$ are the height, the width and the input channels, is fed into both paths. For the transformer path,  $x$ is divided into uniform, non-overlapping patches  $x_v \in \mathbb{R}^{N (S^2 \times C)}$, where ($S\times S\times C$) is the patch size and $N$ is the number of the patches $N = (H\times W) /S^2$. These patches are projected into embedding space $z_0$ using a convolutional kernel $E\in \mathbb{R}^{ (S^2. C)\times K}$, where $K$ is the  dimensionality of the embedding space, which is fixed for all of the transformer layers. $z_0$ is defined by:

\begin{equation}
\label{eq:1}
z_0 = [x_v^1 E;x_v^2 E;.........;x_v^N E]
\end{equation}

\begin{figure*}[htp]
\centering
    \includegraphics[width = 6in, height = 4in] {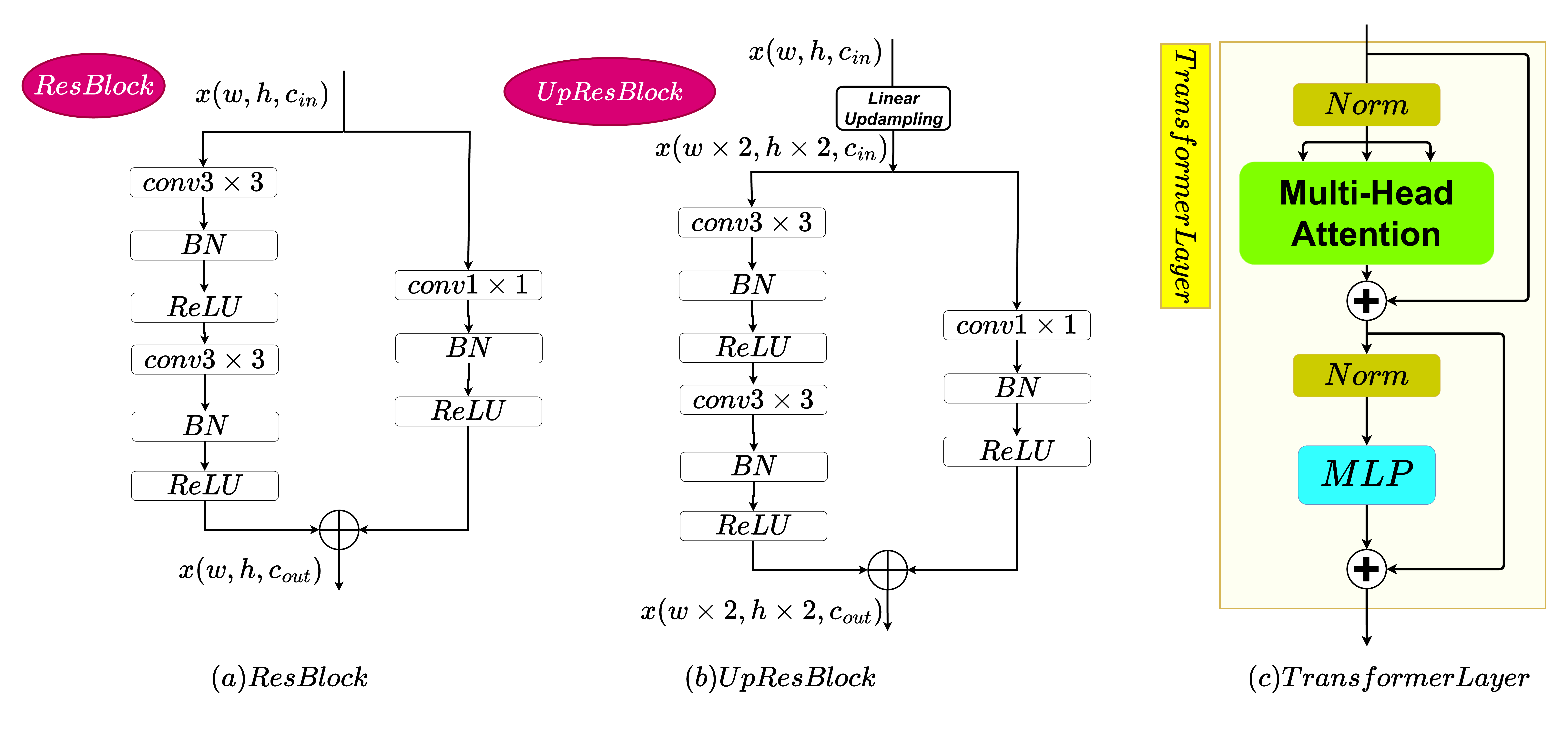} 
    \caption{Description of ResBlock, UpResBlock and TransformerLayer. }
    \label{fig:blocks}
\end{figure*}

The embedded features  $z_0\in \mathbb{R}^{N\times K}$ are fed into Transformer layers similar to \cite{dosovitskiy_image_2020, vaswani_attention_2017}. As shown in Figure \ref{fig:blocks}-c, the Transformer layer consists of two Layernorm (LN) blocks, Multi-Head Self-Attention (MSA) block, a multilayer perceptron (MLP) block and  residual connections.  For the Transformer layer ($l$), the  embedded input features $z_{l-1}$ are fed into  Layernorm (LN), followed by a Multi-Head Self-Attention block, which is then summed with  $z_{l-1}$ by a residual connection, as depicted in equation (\ref{eq:2}):

\begin{equation}
\label{eq:2}
z_l^\prime = MSA(\;LN(z_{l-1})\;) + z_{l-1}
\end{equation}
The embedded features of  $z_{l-1}$ which were passed by the first LN are denoted by  $s = LN(z_{l-1})$. These feature maps are divided into equal features of  $h$ heads, $s = [s_1, s_2, ...., s_h]$, each has $K/h$ dimension features. MSA is defined by:

\begin{equation}
\label{eq:3}
MSA = U_{msa}([SA_1(s_1); SA_2(s_2); ........ ; SA_h(s_h)])
\end{equation}
where $SA_1, SA_2, ... , SA_h$ are Self-Attention Blocks, and $U_{msa} \in \mathbb{R}^{ K\times K}$ is weighting matrix for the SA features. 

The $z_l^\prime$ is fed into  Layernorm (LN) block followed by MLP block then summed with  $z_{l-1}$ by residual connection, as depicted in equation (\ref{eq:4}):

\begin{equation}
\label{eq:4}
z_l = MLP(\;LN(z_l^\prime)\;) + z_l^\prime
\end{equation}
where $MLP$ consists of  two Linear layers with a GELU non-linearity. The first Linear layer ($MLP_1  \in \mathbb{R}^{ K\times K_{MLP}}$) projects $LN(z_l^\prime)$ into $K_{MLP}$, then the second  Linear layer ($MLP_2  \in \mathbb{R}^{ K_{MLP}\times K}$) projects it back to  $K$ features.

In our approach, $L$ is set to $12$, $h$ to $12$, $K = 786$, $K_{MLP}$ = 3072 and S=$16\times 16$ pixels. Consequently,  $W = H = 224$, the number of patches $= 196$.

To obtain diversity of learned features from  different Transformer layers (levels), the embedded features of layers 4, 7, 10 and 12 are selected. These layers are denoted as $Tr_{1}$,  $Tr_{2}$, $Tr_{3}$ and $Tr_{4}$, respectively. Consequently, four layers from the Transformer path were injected into  D-TrAttUnet, all of which have the shape $196\times 786$. To have 3D shape, $z_l$ is reshaped to $ 14\times 14\times 786$, since $14\times 14= 196$. The reshaped features embedded features for $Tr_{1}$,  $Tr_{2}$, $Tr_{3}$ and $Tr_{4}$ are denoted by $z_{1}$,  $z_{2}$, $z_{3}$ and $z_{4}$, respectively. To inject the transformer features into different layers of D-TrAttUnet and combine them with the CNN layers, UpResBlock is introduced as depicted in Figure \ref{fig:blocks}-b.  UpResblock consists of linear upsampling followed by ResBlock  as depicted in Figure  \ref{fig:blocks}-a. ResBlock consists of two 3 by 3 convolutional block, each followed by Batch Normalization and ReLU activation function. In addition,  the input is added with the output of two convolutional layers using the residual connection, which consists of 1 by 1 convolutional block, followed by Batch Normalization and ReLU activation function, as shown in equations (\ref{eq:5}) and (\ref{eq:6}): 

\begin{equation}
\label{eq:5}
x_{out_1} = ReLU(\;BN(Conv3\times3_1(x_{in})\;) 
\end{equation}

\begin{equation}
\label{eq:6}
\begin{split}
x_{out} = ReLU(\;BN(Conv3\times3_2(x_{out_1}))) + \\ ReLU(\;BN(Conv1\times1(x_{in})))
\end{split}
\end{equation}

where $Conv3\times3_1 \in \mathbb{R}^{3\times 3\times C_{out}}$, $Conv3\times3_2 \in \mathbb{R}^{3\times 3\times C_{out}}$ and $Conv1\times1 \in \mathbb{R}^{1\times 1\times C_{out}}$.

\begin{equation}
\label{eq:7}
z_{up_1} = UpResBlock(\; UpResBlock(\;UpResBlock(z_{1r})))
\end{equation}

\begin{equation}
\label{eq:8}
z_{up_2} =  UpResBlock(\;UpResBlock(z_{2r}))
\end{equation}

\begin{equation}
\label{eq:9}
z_{up_3} =UpResBlock(z_{3r})
\end{equation}

\begin{equation}
\label{eq:10}
z_{up_4} = ResBlock(z_{4r})
\end{equation}
Equations (\ref{eq:7}), (\ref{eq:8}) and (\ref{eq:9})  illustrate the number  of UpResBlocks required to match the output of the transformer layers to the CNN path layers, using three, two, and one UpResBlock for $Tr_{1}$,  $Tr_{2}$ and $Tr_{3}$, respectively. For the $Tr_{4}$ layer, ResBlock is used instead of UpResBlock since no upsampling is required here, as depicted in equation (\ref{eq:10}). 

On the other hand, the Encoders Fusion path has five layers which will be denoted by $Un_{1}$,  $Un_{2}$, $Un_{3}$, $Un_{4}$ and $Un_{5}$, respectively. The first layer uses ResBlock on the input image $x \in \mathbb{R}^{H\times W\times C}$ to obtain the first encoder feature maps as shown in equation (\ref{eq:11}).



\begin{equation}
\label{eq:11}
x_{1} = ResBlock(x)
\end{equation}

The second, third, fourth and fifth Encoders Fusion layers combine the transformer features with the max-pooled (MP) features of the previous CNN layer as shown in equations (\ref{eq:12}), (\ref{eq:13}), (\ref{eq:14}) and (\ref{eq:15}):

\begin{equation}
\label{eq:12}
x_{2} =  ResBlock([z_{up_1}, MP(x_{1})])
\end{equation}

\begin{equation}
\label{eq:13}
x_{3} = ResBlock([z_{up_2}, MP(x_{2})])
\end{equation}

\begin{equation}
\label{eq:14}
x_{4} = ResBlock([z_{up_3}, MP(x_{3})])
\end{equation}
\begin{equation}
\label{eq:15}
x_{5} = ResBlock([z_{up_4}, MP(x_{4})])
\end{equation}

\begin{figure*}[htp]
\centering

\includegraphics[width = 6in, height=1.5in] {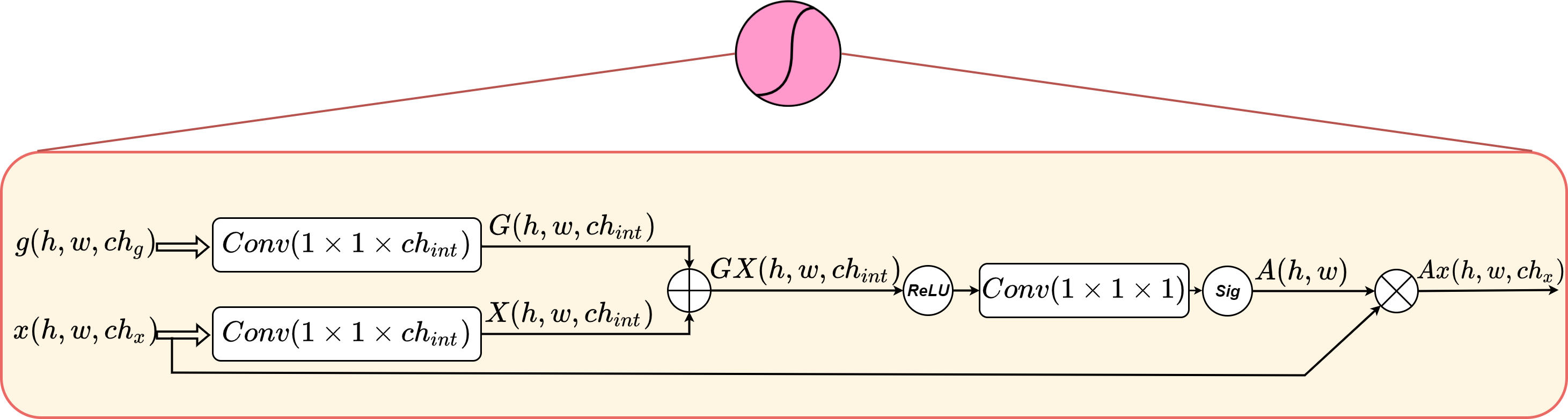}

\label{fig:attgate}
\caption{  Attention Gate block, where g is the gating signal and the x is the input feature maps. $A(h,w)$ is the obtained  spatial attention, which is applied for all channels of the input feature maps (x). }
    \label{fig:att}
\end{figure*}

In our proposed D-TrAttUnet, dual decoders are  used. As shown in Figure \ref{fig:approachA}, the first decoder aims to segment the infection, while, the second one aims to segment the lungs regions. The objective of adding another decoder for lungs segmentation is to make the encoder concentrating inside the lungs, where the infection occurs. Furthermore, it allows to distinguish between lung and non-lung tissues, which can confuse the model, since the non-lung tissues may look similar to the infection, especially if the infection appears as a consolidation. 

The bottleneck feature maps of the $x_{5}$ encoder are fed into the first expansion layer of the two decoders
First, $x_{5}$ is upsampled using a linear transformation to obtain  $d_{5}$, and then passed to the two decoders as shown in equation (\ref{eq:161}). On the other hand, the encoder feature maps $x_{1}, x_{2}, x_{3}$ and  $x_{4}$  are fed to the two decoders layers of D-TrAttUnet through skip connections, as shown in Figure \ref{fig:approachA}.  Following the Att-Unet architecture \cite{oktay_attention_2018}, three linear upsampling layers (US), four decoder layers, four attention gates, and four ResBlocks are used for each decoder, as shown in the following equations:

\begin{equation}
\label{eq:161}
d_{5} = US(x_{5})
\end{equation}

\begin{equation}
\label{eq:16}
d_{41} = ResBlock(\;[AttGate(x_{4}, US(x_{5})); US(x_{5}]))
\end{equation}

\begin{equation}
\label{eq:17}
d_{31} = ResBlock(\;[AttGate(x_{3}, US(d_{41})); US(d_{41}]))
\end{equation}

\begin{equation}
\label{eq:18}
d_{21} = ResBlock(\;[AttGate(x_{2}, US(d_{31})); US(d_{31}]))
\end{equation}
\begin{equation}
\label{eq:19}
d_{11} = ResBlock(\;[AttGate(x_{1}, US(d_{21})); US(d_{21}]))
\end{equation}

Similarly, $d_{42}, d_{32}, d_{22} $and $ d_{12}$ are obtained for the lung segmentation decoder. Finally, two convolutional 1 by 1 layers are used to match the feature map dimension of $d_{11}$ and $d_{12}$ to the infection and lungs masks prediction, which consist of a single channel for the lung and binary segmentation and three channels for the multi-classes segmentation.

The Attention Gate (AG) is depicted in Figure \ref{fig:att} \cite{oktay_attention_2018}, and it is defined as follows:

\begin{equation}
\label{eq:20}
M_{att} = \psi_{i} \; ( ReLU ( BN(\;W_{x} \; x_{i}) + BN(W_{g} \;g_{i})))
\end{equation}

where $W_{x} \in \mathbb{R}^{1\times 1\times C_{int}}$ and $W_{g} \in \mathbb{R}^{1\times 1\times C_{int}}$ are two linear transformations that transform the channels $c_{x}$ and $c_{g}$ from  $x_{i}$ and $g_{i}$, respectively, to $c_{int}$. $ \psi_{i} $ consists of $W_{\psi_{i}} \in \mathbb{R}^{1\times 1\times 1}$ followed by BatchNormalization (BN) and sigmoid activation function to learn the spatial attention coefficient $M_{att_{p}}$ for each pixel. The obtained spatial coefficients $M_{att}$ are applied to the skip feature maps of the encoder  $x_{i}$.

\begin{equation}
\label{eq:21}
x_{att} = M_{att} \;  x_{i}
\end{equation}

\section{Datasets and Evaluation Metrics}
\label{S:4}

\subsection{Datasets}

Three publicly available datasets are used to evaluate the performance of our proposed approach for both binary and multi-classes segmentation tasks. Table \ref{tab:datasets} summarises the datasets used. Dataset\_1 \cite{COVID-19-Dataset} consists of only 100 slices that show Covid-19 infection. These slices has lungs and multi-classes infection (GGO and Consolidation) masks.

Dataset\_2 \cite{COVID-19-Dataset} consists of nine 3D CT-scans. In total, there are 829 slices, of which 373 slices show evidence of Covid-19 infection. This dataset was labelled by expert radiologists, where the masks of lungs, binary infection (non-infected and infected), and multi-classes labels (non-infected, GGO and Consolidation) were provided.

Dataset\_3 \cite{2021MedPh..48.1197M} was constructed using 20 3D CT-scans. In total, Dataset\_3 consists of 3520 slices, of which 1844 slices were labelled as infected by expert radiologists. Both lung regions and binary infection (non-infected and infected) masks were provided for each pixel by the radiologists.

For the binary segmentation task, both Dataset\_2 and Dataset\_3 are divided into 70\%-30\% splits, which are considered as training and testing splits, respectively. For the multi-classes segmentation task, Dataset\_2 and 50\% of Dataset\_1 (50 slices) are used as the training data, the remaining 50 slices of Dataset\_1 are used as the testing data.

\begin{table}
 \caption{The used datasets summary. }
\label{tab:datasets}
\centering
\begin{tabular}{|p{1.2cm}|p{4.2cm}|p{1.1cm}|p{0.9cm}|}
\hline

 \textbf{Name}  & \textbf{Dataset} &\textbf{\#CT-Scans} &   \textbf{\#Slices} \\\hline
Dataset\_1& COVID-19 CT segmentation \cite{COVID-19-Dataset}& 40& 100  \\ \hline 
Dataset\_2&  Segmentation dataset nr. 2 \cite{COVID-19-Dataset}&9 &  829  \\ \hline 
Dataset\_3&COVID-19-CT-Seg dataset \cite{2021MedPh..48.1197M} &20 & 3520 \\ \hline

\end{tabular}
\end{table}

\subsection{Evaluation Metrics}
To evaluate the performance of different approaches, we used three evaluation metrics which are: F1-score (F1-S), Dice-score (D-S), Intersection over Union (IoU).

It should be noted that F1-score and Intersection over Union (IoU) are micro metrics, where they are calculated for all images at one time using True Positives ($TP$),  True Negatives ($TN$), False Positives ($FP$) and  False Negatives ($FN$). However, the Dice-score is the macro version of the $F1-score$. For $N$ training or test images, it is defined by: 

\begin{equation}\label{eq:30}
\text{Dice-score} = 100 \cdot \frac{1}{N} \sum_{i=1}^{N}{ 2 \cdot \frac{TP_i}{2 \cdot TP_i+ FP_i + FN_i }} 
\end{equation}
where $TP_i$, $TN_i$, $FP_i$ and  $FN_i$ are True Positives, True Negatives, False Positives and  False Negative for the $i$th image, respectively.

\section{Experiments and results}
\label{S:5}
\subsection{Experimental~Setup}

To produce our experiments, we mainly used Pytroch \cite{paszke_pytorch_2019} library for deep learning. Each architecture is trained for 60 epochs with an initial learning rate of 0.1 which decays by 0.1 after 30 epochs, followed by another decay of 0.1 after 50 epochs. The batch size is set to 6 images and NVIDIA GPU Device GeForce TITAN RTX 24 GB is used. Three types of active data augmentation are used; random rotate with an angle between $-35^{\circ}$ and $35^{\circ}$ with a probability of 10\% and  random Horizontal  and vertical Flipping with probability of 20\% for each. Adam is the used optimizer and Binary Cross Entropy  and Cross Entropy are used as the losses for the Binary and Multi-classes segmentation tasks, respectively. It should be noted that the loss for the infection segmentation task is weighted by 0.7 and the loss for the lung segmentation is weighted by 0.3. The goal of giving more weight to the loss of the infection task is to have the training focus more on the infection segmentation task.

\subsection{Binary Segmentation}

In this section, we evaluate the performance of the proposed D-TrAttUnet and compare its performance with Unet \cite{ronneberger_u-net_2015}, Att-Unet  \cite{oktay_attention_2018}, Unet++ \cite{zhou_unet_2018}, CopleNet \cite{wang_noise-robust_2020}, AnamNet \cite{paluru_anam-net_2021}, and SCOATNet \cite{zhao2021scoat}, it should be noted that each experiment was repeated five times. The results shown represent the average of the best result based on the F1-score on the validation data $\pm$ the standard deviation of the five runs.

\begin{table}
 \caption{ Performance evaluation of the proposed D-TrAttUnet and Unet \cite{ronneberger_u-net_2015}, Att-Unet  \cite{oktay_attention_2018}, Unet++ \cite{zhou_unet_2018}, CopleNet \cite{wang_noise-robust_2020}, AnamNet \cite{paluru_anam-net_2021}, and SCOATNet \cite{zhao2021scoat} on Dataset\_2}
\label{tab:dataset9}
\centering
\begin{tabular}{|p{2cm}|p{1.8cm}|p{1.8cm}|p{1.8cm}|}
\hline

 \textbf{Model}  & \textbf{F1-S} &\textbf{D-S} &   \textbf{IoU} \\
\hline

Unet  & 47.36$\pm$14.54 &22.23$\pm$6.51& 32.24$\pm$12.76 \\\hline

Att-Unet  & 50.61$\pm$12.41 & 23.83$\pm$5.16  &  34.82$\pm$11.45  \\\hline

UNet++  & 55.20$\pm$12.14& 27.01$\pm$5.75& 39.05$\pm$10.96  \\\hline \hline


CopleNet  & 60.92$\pm$9.16&  26.09$\pm$4.11 &  44.42$\pm$9.44 \\\hline

AnamNet  &38.87$\pm$3.8& 20.13$\pm$1.66&  27.20$\pm$2.91 \\\hline

SCOATNet  &45.28$\pm$18.46 & 19.87$\pm$7.52&  31.12$\pm$15.56\\\hline 

\hline

\bf{D-TrAttUnet}& \bf{74.44$\pm$2.38}& \bf{36.86$\pm$2.63}&  \bf{ 59.34$\pm$3.01} \\\hline
\end{tabular}

\end{table}

\begin{table}
\caption{ Performance evaluation of the proposed D-TrAttUnet and Unet \cite{ronneberger_u-net_2015}, Att-Unet  \cite{oktay_attention_2018}, Unet++ \cite{zhou_unet_2018}, CopleNet \cite{wang_noise-robust_2020}, AnamNet \cite{paluru_anam-net_2021}, and SCOATNet \cite{zhao2021scoat} on Dataset\_3.}
\label{tab:dataset2}
\centering
\begin{tabular}{|p{2cm}|p{1.8cm}|p{1.8cm}|p{1.8cm}|}
\hline

 \textbf{Model}  & \textbf{F1-S} &\textbf{D-S} &   \textbf{IoU} \\
\hline

Unet  & 67.72$\pm$2.65 &  36.14$\pm$ 1.31& 51.25 $\pm$ 3 \\\hline

Att-Unet  & 62.85$\pm$7.06 & 33.51$\pm$2.5 & 46.19 $\pm$7.03 \\\hline

UNet++  &  66.10 $\pm$3.81&  36.66$\pm$1.97&  49.49$\pm$4.28 \\\hline \hline


CopleNet  & 63.55$\pm$7.57&  34.56$\pm$3.52 &  47.02$\pm$8.01 \\\hline

AnamNet  &69.34$\pm$3.33 & 37.66$\pm$1.81 & 53.17$\pm$3.79 \\\hline

SCOATNet  & 70.27$\pm$2.39 & 38.47$\pm$0.56&  54.22$\pm$2.86 \\\hline 
\hline


\bf{D-TrAttUnet}& \bf{75.42 $\pm$ 0.97}& \bf{ 40.57 $\pm$ 0.28}&  \bf{60.55 $\pm$ 1.24} \\\hline
\end{tabular}

\end{table}

\subsection{Multi-classes Segmentation}
Table \ref{tab:MC} summarizes the obtained results of our proposed D-TrAttUnet architecture and the comparison methods for multi-classes Covid-19 segmentation. For GGO, our approach outperforms the comparison architectures, where it is noticed that many of the comparison architectures achieved close results. Our architecture achieves better result than the best architecture for each metric; 1.44\% for $F_1$-score (Unet), 1.85\% for Dice-score (AnamNet), and 1.58\% for IoU (SCOATNET). Similarly, our approach surpasses the comparison methods for the consolidation segmentation. In more details, the proposed D-TrAttUnet architecture outperforms the best comparison architecture (which is SCOATNET) by 11.72\%, 7.59\% and 10.33\% for $F_1$-score, Dice-score, and IoU, respectively. These results show that our approach has a good capability to deal with unbalanced classes, which corresponds to the real scenario of Covid-19 infection, where GGO infection type is usually more frequent than consolidation.

\begin{table*}[ht!]
 \caption{Performance evaluation of our proposed approach (D-TrAttUnet), Unet \cite{ronneberger_u-net_2015}, Att-Unet  \cite{oktay_attention_2018}, Unet++ \cite{zhou_unet_2018}, CopleNet \cite{wang_noise-robust_2020}, AnamNet \cite{paluru_anam-net_2021}, and SCOATNet \cite{zhao2021scoat} for multi-classes Covid-19 segmentation (No-infection, GGO and Consolidation).  }
 \begin{center}
\label{tab:MC}
\centering
\begin{tabular}{|c|l|c|c|c|c|c|c|}

\hline
 Ex &{\multirow{2}{*}    \textbf{Architecture}}    & \multicolumn{3}{|c|}{\textbf{GGO}}& \multicolumn{3}{|c|}{\textbf{Consolidation}} \\
\cline{3-8}


& & \textbf{F1-S} &\textbf{D-S} &   \textbf{IoU}& \textbf{F1-S} &\textbf{D-S} &   \textbf{IoU}  \\
\hline

1&Unet   & 65.81$\pm$1.26 & 50.13$\pm$1.31 & 49.06$\pm$1.41 & 31.35$\pm$12.96 & 15.45$\pm$5.66 & 19.26$\pm$8.76 \\\hline

2&Att-Unet & 64.81$\pm$1.89 & 50.44$\pm$1.35 & 47.97$\pm$2.06 & 39.04$\pm$6.81 & 19.26$\pm$3.55 & 24.48$\pm$5.31 \\\hline
2&Unet++ & 65.69$\pm$1.29 & 51.65$\pm$4.12 & 48.92$\pm$14.2 & 31.31$\pm$6.67 & 16.86$\pm$4.48 & 18.75$\pm$4.73 \\\hline \hline

4&CopleNet & 60.44$\pm$1.54 & 46.25$\pm$3.13 & 43.33$\pm$1.61 & 29.70$\pm$10.29 & 16.46$\pm$4.76 & 17.90$\pm$7.52 \\\hline 
5&AnamNet & 65.10$\pm$ 3.56& 51.69$\pm$4.81 & 48.36$\pm$3.82 & 31.97$\pm$6.12 & 18.06$\pm$4.61 & 19.18$\pm$4.36 \\\hline 
6&SCOATNET  & 65.77$\pm$3.28 & 50.80$\pm$4.63 & 49.09$\pm$3.56 & 43.52$\pm$1.67 & 23.32$\pm$2.07 & 27.83$\pm$1.38\\\hline \hline

7&\bf{D-TrAttUnet}& \bf{67.25$\pm$1.40} & \bf{53.54$\pm$ 1.24}& \bf{50.67$\pm$1.59}& \bf{55.24$\pm$0.97}& \bf{30.91$\pm$1.67} & \bf{38.16$\pm$0.93} \\\hline 
\end{tabular}
\end{center}
\end{table*}


\subsection{Ablation Study}

\begin{table}[ht!]
 \caption{Ablation study of Binary Segmentation scenario. The experimental results of Dataset\_2 and Dataset\_3  are summarized with investigating the effectiveness of the following components: Attention Gate (AG), Dual-Decoder (DD) and Transformer Encoder (TrEc).  }
 \begin{center}
\label{tab:abbinary1}
\centering
\begin{tabular}{|l|l|ccc|c|c|c|}

\hline
 \multirow{2}{*}\textbf{Ex} & { \multirow{2}{*}   \textbf{Architecture}}   &\multicolumn{3}{|c|}{\textbf{Ablation}}  & \multicolumn{3}{|c|}{\textbf{Dataset\_2}} \\
\cline{3-8}

& &\textbf{AG} &\textbf{DD} &\textbf{TrEc}       & \textbf{F1-S} &\textbf{D-S} &   \textbf{IoU} \\
\hline

1&Unet (baseline)& \xmark & \xmark & \xmark & 47.36$\pm$14.54 &22.23$\pm$ 6.51& 32.24$\pm$12.76  \\\hline 

2&AttUnet  (baseline)& \cmark & \xmark & \xmark & 50.61$\pm$12.41 & 23.83$\pm$5.16  &  34.82$\pm$11.45 \\\hline 

3&D-TrUnet&\xmark & \cmark & \cmark  & 70.37$\pm$ 3.98 & 36.46$\pm$2.56 & 54.43$\pm$4.80  \\\hline

4&D-AttUnet&\cmark & \cmark & \xmark   &63.43$\pm$ 6.35 & 30.39$\pm$4.08 & 46.76$\pm$6.81 \\\hline
5&TrAttUnet&\cmark & \xmark & \cmark  & 67.33$\pm$6.72 & 32.52$\pm$3.46 & 51.14$\pm$7.76   \\\hline


6&\bf{D-TrAttUnet}&\cmark & \cmark & \cmark & \bf{74.44 $\pm$2.38}& \bf{36.86$\pm$2.63}&  \bf{59.34$\pm$ 3.01} \\\hline

\end{tabular}
\end{center}
\end{table}

\begin{table}[ht!]
 \caption{Ablation study of Binary Segmentation scenario. The experimental results of Dataset\_2 and Dataset\_3  are summarized with investigating the effectiveness of the following components: Attention Gate (AG), Dual-Decoder (DD) and Transformer Encoder (TrEc).  }
 \begin{center}
\label{tab:abbinary2}
\centering
\begin{tabular}{|l|l|ccc|c|c|c|}

\hline
 \multirow{2}{*}\textbf{Ex} & { \multirow{2}{*}   \textbf{Architecture}}   &\multicolumn{3}{|c|}{\textbf{Ablation}} & \multicolumn{3}{|c|}{\textbf{Dataset\_3}} \\
\cline{3-8}

& &\textbf{AG} &\textbf{DD} &\textbf{TrEc}       & \textbf{F1-S} &\textbf{D-S} &   \textbf{IoU}  \\
\hline

1&Unet (baseline)& \xmark & \xmark & \xmark  & 67.72$\pm$ 2.65 &  36.14$\pm$ 1.31& 51.25$\pm$ 3.01 \\\hline 

2&AttUnet  (baseline)& \cmark & \xmark & \xmark &   62.85$\pm$7.06 & 33.51$\pm$2.5 & 46.19 $\pm$7.03 \\\hline 

3&D-TrUnet&\xmark & \cmark & \cmark   & 74.43$\pm$0.51 & 40.06$\pm$0.35 & 59.27$\pm$0.65  \\\hline

4&D-AttUnet&\cmark & \cmark & \xmark   & 72.21$\pm$0.53 & 38.88$\pm$ 0.79&  56.51$\pm$0.64 \\\hline
5&TrAttUnet&\cmark & \xmark & \cmark   & 70.57$\pm$2.41 & 39.05$\pm$0.8 & 54.58$\pm$2.86  \\\hline


6&\bf{D-TrAttUnet}&\cmark & \cmark & \cmark &\bf{75.42$\pm$0.97}&  \bf{40.57$\pm$0.28} & \bf{60.55$\pm$1.24} \\\hline

\end{tabular}
\end{center}
\end{table}

The aim of this section is to examine the significance of each element of the proposed D-TrAttUnet architecture. Tables \ref{tab:abbinary1} \ref{tab:abbinary2} summarize the results on Dataset\_2 and Dataset\_3 for the binary segmentation, respectively. In this table, we study the importance of the Attention Gate (AG), the Dual Decoders (DD) and the Transformer Encoder (TrEc). From Experiments 1 and 2, it is noticed that the attention gate improves the results of Unet architecture in Dataset\_2.  In contrast, in Dataset\_3, Unet's results fell when the Attention Gate was added. This raises the question about the feasibility of using the attention gate for Covid-19 segmentation. To answer this question, we removed the attention gate from our approach in Experiment 3. By comparing the results of Experiment 3 and our proposed D-TrAttUnet architecture, we notice that the attention gate is very essential component in our approach, as the results on Dataset\_2 improve by 4.07\%,  0.4\%, and 4.91\%  for $F_1$-score, Dice-score, and IoU, respectively. Similarly, the results on Dataset\_3 improve by 1\%,  0.5\%, and 1.28\%  for $F_1$-score, Dice-score, and IoU, respectively.The integration of the transformer layers in the encoding phase provides richer feature extraction, which is passed to the attention gate via the skip connections. The attention gates with richer features are properly activated to select the most important features from the encoder and the upsampled features of the previous decoder layer.  

Experiment 4 ( in Tables \ref{tab:abbinary1} \ref{tab:abbinary2}) depicts the results obtained  when the transformer encoder is removed. From these results, it can be seen that without the transformer encoder, the results decreased, especially for the Dataset\_2, which consists of only 9 CT-scans. In more details, the transformer encoder improves the results on Dataset\_2 by 11\%,  6.47\%, and 12.58\%  for $F_1$-score, Dice-score, and IoU, respectively. Similarly, the results of Dataset\_3 are improved by 3.21\%,  1.69\%, and 4.04\%  for $F_1$-score, Dice-score, and IoU, respectively. This proves the efficiency of combining the transformer  and convolutional layers as encoder, especially when available data are limited, as is the case for pandemics. 

The comparison between the results of experiment 5 and 6 (Tables \ref{tab:abbinary1} \ref{tab:abbinary2}) shows the importance of using Dual-Decoders. For both Dataset\_2 and Dataset\_3,  the results are improved by adding the second decoder for lung segmentation at the same time as infection segmentation.

\begin{table}[ht!]
 \caption{Ablation study of Binary Segmentation scenario. The experimental results using Dataset\_1 and Dataset\_2  are summarized with investigating the effectiveness of the following components: Attention Gate (AG), Dual-Decoder (DD) and Transformer Encoder (TrEc).  }
 \begin{center}
\label{tab:abMC}
\resizebox{1\linewidth}{!}{\centering
\begin{tabular}{|l|l|ccc|c|c|c|c|c|c|}

\hline
  \multirow{2}{*}\textbf{Ex} & {\multirow{2}{*}    \textbf{Architecture}}   &\multicolumn{3}{|c|}{\textbf{Ablation}}  & \multicolumn{3}{|c|}{\textbf{GGO}}& \multicolumn{3}{|c|}{\textbf{Consolidation}} \\
\cline{3-11}

& &\textbf{AG}&\textbf{DD} &\textbf{TrEc}       & \textbf{F1-S} &\textbf{D-S} &   \textbf{IoU}& \textbf{F1-S} &\textbf{D-S}&   \textbf{IoU}  \\
\hline

1&Unet  (baseline)& \xmark& \xmark & \xmark & 65.81$\pm$1.26 & 50.13$\pm$1.31 & 49.06$\pm$1.41 & 31.35$\pm$12.96 & 15.45$\pm$5.66 & 19.26$\pm$8.76\\\hline
2&AttUnet  (baseline)& \cmark& \xmark & \xmark & 64.81$\pm$1.89 & 50.44$\pm$1.35 & 47.97$\pm$2.06 & 39.04$\pm$6.81 & 19.26$\pm$3.55 & 24.48$\pm$5.31\\\hline

3&D-TrUnet& \xmark&\cmark & \cmark & 63.77$\pm$1.69 & 49.80$\pm$2.97 & 46.83$\pm$1.82& 50.39$\pm$2.08& 29.19$\pm$4.55 & 33.71$\pm$1.84 \\\hline
4&D-AttUnet& \cmark&\cmark & \xmark & 65.20$\pm$0.95&  51.62$\pm$1.42& 48.38$\pm$1.05 & 51.15$\pm$2.16&29.02$\pm$1.40 &  34.39$\pm$1.98 \\\hline
5&TrAttUnet& \cmark&\xmark & \cmark & 65.69$\pm$1.29&  51.65$\pm$4.12& 48.92$\pm$1.42 & 48.15$\pm$1.75&27.23$\pm$4.52 &  31.73$\pm$1.50 \\\hline

 \bf{6}&\bf{D-TrAttUnet}& \cmark&\cmark & \cmark & \bf{67.25$\pm$1.40} & \bf{53.54$\pm$ 1.24}& \bf{50.67$\pm$1.59}& \bf{55.24$\pm$0.97}& \bf{30.91$\pm$1.67} & \bf{38.16$\pm$0.93} \\\hline

\end{tabular}}
\end{center}
\end{table}

Table \ref{tab:abMC} depicts the ablation experiments for multi-classes segmentation. Similar to the ablation experiments of the binary segmentation, the importance of Attention Gate (AG), Dual Decoders (DD) and Transformer Encoder  (TrEc) are studied in Table \ref{tab:abMC}. From the results of  Unet and AttUnet (Experiments 1 and 2), it can be observed that the attention gate is very useful for Consolidation segmentation, where the $F_1$-score is improved by 8.7\%. However, for GGO segmentation,
adding the Attention Gate decreases the performance a little for $F_1$-score and IoU. 
On the other hand, the Attention Gate turns out to be a very important component for both classes of segmentation (GGO and Consolidation) for our approach. More specifically, Attention Gate significantly improves the segmentation results of GGO by
 by 3.48\%, 3.74\%, and 3.84\%  for $F_1$-score, Dice-score, and IoU, respectively. Similarly, the results of consolidation are improved by and 8.41\%,  1.72\%, and 4.45\%  for $F_1$-score, Dice-score, and IoU, respectively. Similar to what found in the binary segmentation, the Attention Gate is also very important in multi-classes segmentation because it plays a crucial role in identifying the most important features from the encoder layers. Especially, that the multi-classes segmentation is a very complicated task compared to the binary segmentation task.  

From the fourth and fifth rows of Table \ref{tab:abMC}, it is very clear that both the Dual-Decoders and Transformer encoder are very important components for  our proposed D-TrAttUnet architecture. 
The results are significantly improved by adding each of them.
Where the results are  considerable ameliorated by adding each element. The improvement is very significant in the consolidation segmentation, adding the Dual-Decoders and Transformer Encoder improves the $F_1$-score results by about 4\% and 6\%, respectively. The ablation study in Tables\ref{tab:abbinary1} \ref{tab:abbinary2}, \ref{tab:abMC} proves the importance of each proposed component for binary and multi-classes Covid-19 segmentation for our approach.


\begin{figure*}[htbp]
\setlength\tabcolsep{1pt}
\settowidth\rotheadsize{Radcliffe Cam}
\begin{tabularx}{\linewidth}{Xp{2pt}Xp{2pt}Xp{2pt}Xp{2pt}Xp{2pt}Xp{2pt}Xp{2pt}X }
 
       \includegraphics[width = 2.35cm,valign=m]   {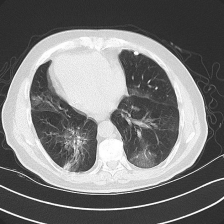} &  
        &  \includegraphics[width = 2.35cm,valign=m] {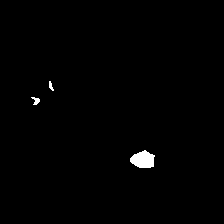}& 
        & \includegraphics[width = 2.35cm,valign=m] {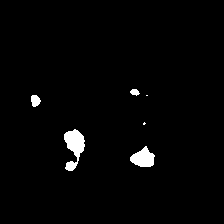}&
        &   \includegraphics[width = 2.35cm,valign=m] {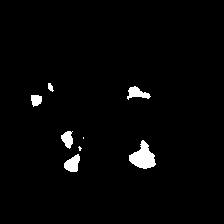}& & \includegraphics[width = 2.35cm,valign=m]{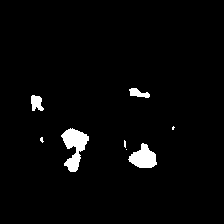}&&
        \includegraphics[width = 2.35cm,valign=m]{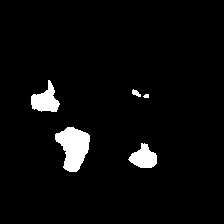}&& \includegraphics[width = 2.35cm,valign=m]{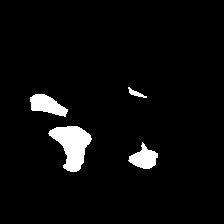}
        
 \\ 
 
       \includegraphics[width = 2.35cm,valign=m]   {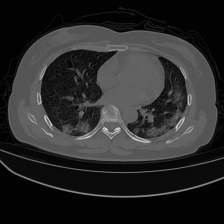} &  
        &  \includegraphics[width = 2.35cm,valign=m] {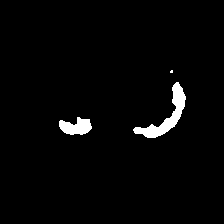}& 
        & \includegraphics[width = 2.35cm,valign=m] {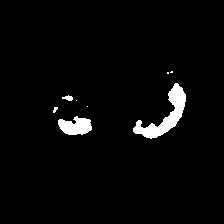}&
        &   \includegraphics[width = 2.35cm,valign=m] {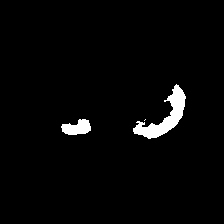}& & \includegraphics[width = 2.35cm,valign=m]{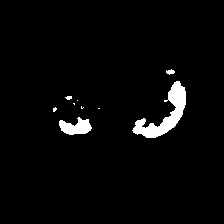}&&
        \includegraphics[width = 2.35cm,valign=m]{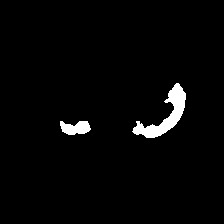}&& \includegraphics[width = 2.35cm,valign=m]{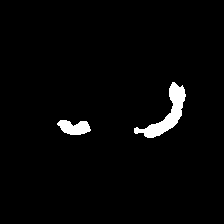}
        
 \\ 
 
       \includegraphics[width = 2.35cm,valign=m]   {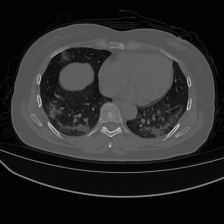} &  
        &  \includegraphics[width = 2.35cm,valign=m] {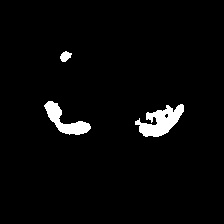}& 
        & \includegraphics[width = 2.35cm,valign=m] {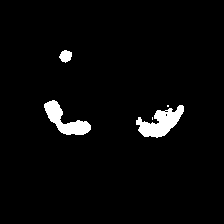}&
        &   \includegraphics[width = 2.35cm,valign=m] {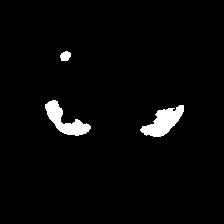}& & \includegraphics[width = 2.35cm,valign=m]{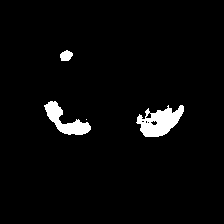}&&
        \includegraphics[width = 2.35cm,valign=m]{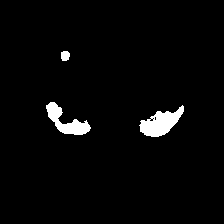}&& \includegraphics[width = 2.35cm,valign=m]{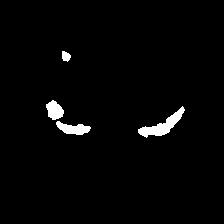}        
 \\  \addlinespace[2pt]
 
       \includegraphics[width = 2.35cm,valign=m]   {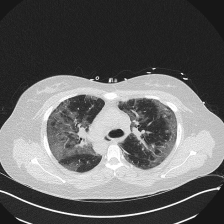} &  
        &  \includegraphics[width = 2.35cm,valign=m] {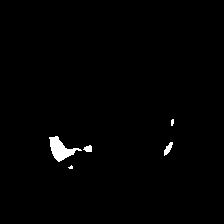}& 
        & \includegraphics[width = 2.35cm,valign=m] {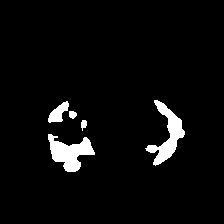}&
        &   \includegraphics[width = 2.35cm,valign=m] {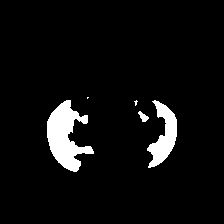}& & \includegraphics[width = 2.35cm,valign=m]{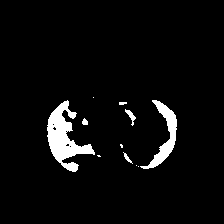}&&
        \includegraphics[width = 2.35cm,valign=m]{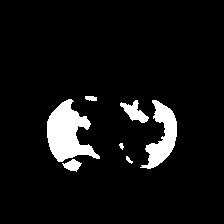}&& \includegraphics[width = 2.35cm,valign=m]{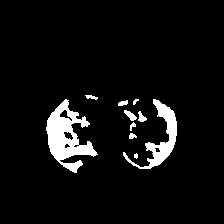}

        \\
        \centering\bf{Slice} &&\centering\bf{AttUnet}&&\centering \bf{Unet++} &&\centering\bf{AnamNet}&&\centering\bf{SCOATNet}&&\centering\bf{D-TrAttUnet}&&\centering\bf{GT}
                      
\end{tabularx}
\caption{Visual comparison of a segmentation model trained with different segmentation architectures for Binary Covid-19 segmentation using Dataset\_2 and Dataset\_3.  }
\label{fig:compvis}
\end{figure*}

\begin{figure*}[htbp]
\setlength\tabcolsep{1pt}
\settowidth\rotheadsize{Radcliffe Cam}
\begin{tabularx}{\linewidth}{Xp{2pt}Xp{2pt}Xp{2pt}Xp{2pt}Xp{2pt}Xp{2pt}Xp{2pt}X }

       \includegraphics[width = 2.35cm,valign=m]   {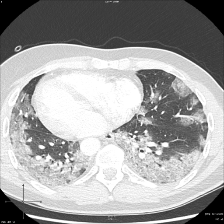} &  
        &  \includegraphics[width = 2.35cm,valign=m] {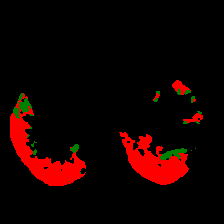}& 
        & \includegraphics[width = 2.35cm,valign=m] {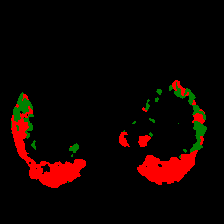}&
        &   \includegraphics[width = 2.35cm,valign=m] {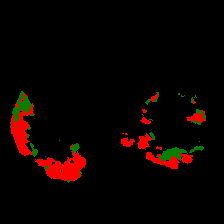}& & \includegraphics[width = 2.35cm,valign=m]{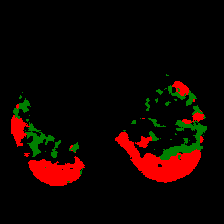}&&
        \includegraphics[width = 2.35cm,valign=m]{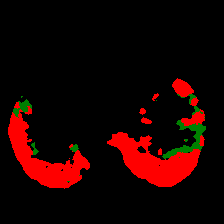}&& \includegraphics[width = 2.35cm,valign=m]{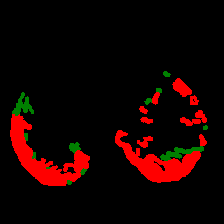}
        
 \\  \addlinespace[2pt]
 
       \includegraphics[width = 2.35cm,valign=m]   {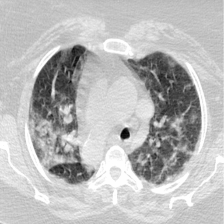} &  
        &  \includegraphics[width = 2.35cm,valign=m] {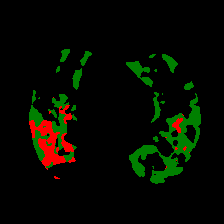}& 
        & \includegraphics[width = 2.35cm,valign=m] {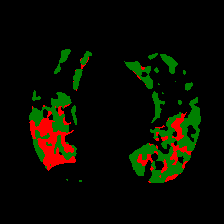}&
        &   \includegraphics[width = 2.35cm,valign=m] {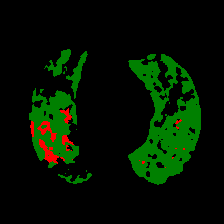}& & \includegraphics[width = 2.35cm,valign=m]{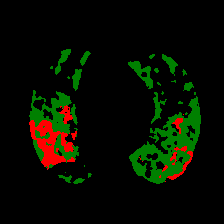}&&
        \includegraphics[width = 2.35cm,valign=m]{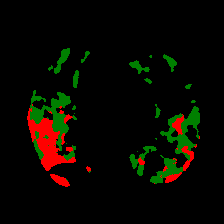}&& \includegraphics[width = 2.35cm,valign=m]{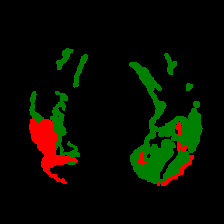}  
        
 \\  
 
       \includegraphics[width = 2.35cm,valign=m]   {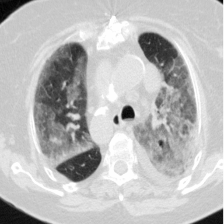} &  
        &  \includegraphics[width = 2.35cm,valign=m] {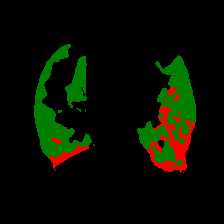}& 
        & \includegraphics[width = 2.35cm,valign=m] {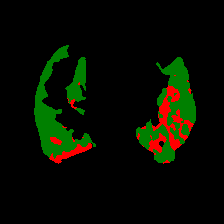}&
        &   \includegraphics[width = 2.35cm,valign=m] {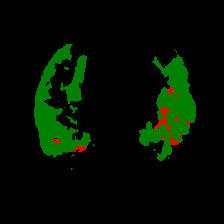}& & \includegraphics[width =2.35 cm,valign=m]{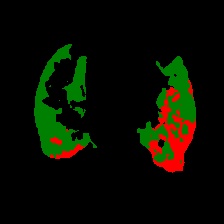}&&
        \includegraphics[width = 2.35cm,valign=m]{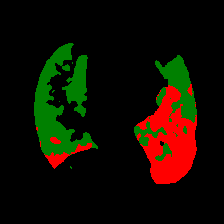}&& \includegraphics[width = 2.35cm,valign=m]{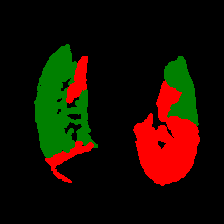}

        \\
        \centering\bf{Slice} &&\centering\bf{AttUnet}&&\centering \bf{Unet++} &&\centering\bf{AnamNet}&&\centering\bf{SCOATNet}&&\centering\bf{D-TrAttUnet}&&\centering\bf{GT}
                      
\end{tabularx}
\caption{Visual comparison of a segmentation model trained with different segmentation architectures for Multi-classes (No-infection, GGO and Consolidation) Covid-19 infection segmentation using Dataset\_1 and Dataset\_2. GGO is presented by the Green color and Consolidation by the red color.  }
\label{fig:compvismc}
\end{figure*}

\section{Discussion}
\label{S:6}

In addition to comparing with state-of-the-art architectures, we visualize some predicted masks for both binary and multi-classes segmentation using our approach and the comparison methods, as shown in Figures \ref{fig:compvis} and \ref{fig:compvismc}. The comparison methods are: Att-Unet  \cite{oktay_attention_2018}, Unet++ \cite{zhou_unet_2018}, AnamNet \cite{paluru_anam-net_2021} and SCOATNet \cite{zhao2021scoat}, which showed a competitive performance with our proposed approach (see Section \ref{S:5}). 

The four visualized examples in Figure \ref{fig:compvis} are from the binary segmentation experiments of Dataset\_2 and Dataset\_3.  The first example shows a case in which infection has spread to both lungs and appears as a GGO and small consolidation region at the bottom of the right lung. The comparison between the AttUnet mask and the ground truth (GT) shows that the AttUnet architecture fails in segmenting most of the infection regions. The  Unet++, AnamNet and SCOATNet masks show improved segmentation performance compared to AttUnet. However,  these architectures still miss some infected regions or segment lung tissues as infection instead. The mask of our proposed approach shows high similarity with  GT in term of the number of regions and their global shape. Both examples 2 and 3 are cases where the infection has a peripheral distribution. The visualized masks show that the proposed D-TrAttUnet is the best architecture consistent with the ground truth. The fourth example depicts a severe case where the infection has spread to most of the lung regions.. The visualized masks  exhibit that our proposed architecture  performs better than the comparison architectures. 

Figure \ref{fig:compvismc} consists of the visualization of three examples masks using our approach and the comparison architectures for multi-classes Covid-19 segmentation. The first example shows a mixture case of GGO and Consolidation, where most of the infected regions consist of consolidation and small GGO regions are attached to the consolidation regions. Unlike the masks of the comparison architectures, the mask of our approach has a high similarity to the ground truth mask for both the consolidation and GGO classes. The second and  third examples also represent a case where both GGO and consolidation are present in both lungs. The infected regions with consolidation are mainly in the lower lobes of both lungs and GGO spreads in both lungs with peripheral and posterior distribution. 
The masks of these examples confirm the observation in the first example, as the predicted masks of D-TrAttUnet show a high similarity to the GT masks for both infection types GGO and consolidation

\section{Conclusion}
\label{S:7}
In this paper, we proposed a CNN-Transformer based approach to segment Binary and Multi-classes Covid-19 infection from CT-scans. Our proposed D-TrAttUnet Encoder merges Transformer and CNN layers to extract richer local, global and  long-range dependency features for Covid-19 segmentation. In addition,  our proposed D-TrAttUnet architecture consists of Dual-Decoder, each one consists of attention gates,  linear Upsampling and Convolutional blocks. The Dual-Decoders are used to segment the infection and the lung regions simultaneously.

To evaluate the performance of our approach, both Binary and multi-classes Covid-19 infection segmentation scenarios were investigated. The proposed D-TrAttUnet architecture outperformed three baseline architectures  and three state-of-the-art architectures on three public Covid-19 segmentation datatsets. The experimental results showed the importance of using attention gates with the compound CNN-Transformer encoder compared to a CNN-only Encoder. The ablation study showed the importance of each component of our proposed approach. Moreover, the combination of all proposed elements leads to better performance with stable and consistent results in both evaluated tasks (Binary and Multi-classes). As future work, we plan to evaluate the performance of our approach in similar segmentation tasks in medical imaging field.


\end{document}